\documentclass[preprint,showpacs,preprintnumbers]{revtex4}
\usepackage{amssymb}
\usepackage{amsmath}
\usepackage{graphicx}
\usepackage{dcolumn}
\usepackage{verbatim}
\usepackage{bm}
\usepackage{color}

\begin{document}

\title{Quantum Fidelity and Thermal Phase Transitions}
\author{H. T. Quan}
\author{F. M. Cucchietti}
\affiliation{Theoretical Division, MS B213, Los Alamos National Laboratory, Los Alamos,
NM, 87545, U.S.A.}
\date{\today}

\begin{abstract}
We study the quantum fidelity approach to characterize thermal phase transitions. 
Specifically, we focus on the mixed-state fidelity induced by a perturbation in temperature.
We consider the behavior of fidelity in two types of second-order thermal phase transitions
(based on the type of non-analiticity of free energy), and we
find that usual fidelity criteria for identifying critical points 
is more applicable to the case of $\lambda$ transitions (divergent second derivatives of free energy).
Our study also reveals limitations of the fidelity approach: 
sensitivity to high temperature thermal fluctuations that wash out
information about the transition, and inability of fidelity to distinguish between 
crossovers and proper phase transitions. In spite of these limitations, however, 
we find that fidelity remains a good pre-criterion for testing thermal phase transitions, 
which we use to analyze the non-zero temperature phase diagram of the 
Lipkin-Meshkov-Glick model.
\end{abstract}

\maketitle

\section{Introduction}
Quantum phase transitions (QPTs) \cite{sachdev}, 
the sudden change in the properties of a quantum many-body system as a control parameter is varied,
have received considerable attention in the past decade. 
Once almost exclusively the domain of condensed matter physics,
the field of quantum critical phenomena has recently 
attracted the attention of the quantum information community: 
some quantum entanglement measurements \cite{rmp07}
such as concurrence \cite{concurrence},
entanglement entropy \cite{kitaev}, and geometric phase \cite{geometric} 
can exhibit singular behavior at quantum critical points. Thus, they can be used 
in place of macroscopic thermodynamic
quantities in classical statistical mechanics -- e.g. specific heat and magnetic susceptibility -- not only
to characterize different QPTs, but also to gain insight on the nature of the quantum critical behavior.

Motivated by the sensitivity to perturbations of quantum systems near a
critical region, one of us and collaborators \cite {quanprl1} proposed to use the
Loschmidt echo \cite{fernando} as another quantum information probe of QPTs.
Based on this work, Zanardi {\em et al} further proposed a geometric measure: the quantum 
fidelity \cite{fidelity} (the overlap) between two ground states 
corresponding to slightly different values of the controlling parameters.
A flurry of work ensued \cite{fidelitystudy}, showing that, despite its simplicity, quantum fidelity 
does indeed capture the dramatic changes in the structure of the ground state at a
quantum critical point. In particular, it has been observed that for second order QPTs
fidelity presents a minimum at the critical point \cite{fidelity}, which became the standard criterion for
detecting quantum criticality with fidelity. Though fidelity is used to study
QPTs at zero temperature, its finite-temperature (thermal state) extension has also been considered \cite {quan7}.  
The motivation behind this approach is similar to that for QPTs: 
The proximity to criticality must be reflected in the geometric distance between
two states separated by a small perturbation 
(either in temperature or in an external parameter).
The fidelity of mixed-states \cite{uhlmann} at finite temperature also gives useful
information about the zero-temperature phase diagram \cite{quan7}.
Studies of finite temperature transitions
using this fidelity approach have been reported for
specific models, such as the Stoner-Hubbard itinerant electron model of magnetism, the BCS model \cite{portugal},
and also the crossover at finite temperature in the one-dimensional transverse Ising model (TIM)  \cite{zanardi1}.
Nevertheless, we find that, in the existing literature, the mechanism for which fidelity
can be used to characterize {\em thermal} phase transitions has not been studied systematically.
In this paper we will study the mixed-state fidelity approach in general second-order
 thermal phase transitions, and explore its applicability and limitations. 
We will focus on {\em non-zero} temperature phase diagrams 
and illustrate our arguments with specific examples.
Finally, we will also discuss the quantum-classical transition of the system when increasing the temperature from a new angle: the relation between quantum fidelity and magnetic susceptibility.
In the rest of this work, and unless explicitly stated, 
we will use the term fidelity to mean mixed-state fidelity. 

This paper is organized as follows: In Section II, we introduce the 
finite-temperature mixed-state fidelity and study its relation 
to the analyticity of free energy.
In particular, we show why fidelity can signal phase transitions by
establishing its relationship to specific heat and magnetic susceptibility.
In Section III we study examples of
two types of second-order thermal phase transitions -- either a divergence or a discontinuity of 
specific heat at critical points--, and discuss the corresponding behavior of mixed-state fidelity. 
The problems in characterizing the second type of transitions with fidelity will be shown.
In Section IV we discuss other limitations of the fidelity approach:
fidelity cannot distinguish between phase transitions and crossovers, 
and at high temperatures thermal fluctuations reduce the effectivity of fidelity for picking out critical points.
In Sec V, we use the Lipkin-Meshkov-Glick model as 
an example to demonstrate that, despite its limitations,
fidelity remains a useful pre-criteria for thermal phase transitions due to its simple form. 

\section{Finite-temperature fidelity and its relation to specific heat and magnetic susceptibility}
The mixed-state fidelity of two thermal states with small perturbations
in temperature and controlling parameter is defined as \cite{uhlmann,quan7}
\begin{equation}
 \mathcal{F}(\beta_0,\lambda_0;\beta_1,\lambda_1)=\mathrm{Tr}\sqrt{\sqrt{\rho_0}\rho_1\sqrt{\rho_0}},\label{1}
\end{equation}
where the thermal states are written in terms of the Hamiltonian $H$ of the system
\begin{equation}
\rho_{\alpha}=\frac{e^{-\beta_{\alpha}H(\lambda_{\alpha})} } {Z(\beta_{\alpha}, \lambda_{\alpha})},
\end{equation}
with the partition function
\begin{equation}
Z(\beta_{\alpha}, \lambda_{\alpha})=\mathrm{Tr}e^{-\beta_{\alpha}H(\lambda_{\alpha})},
\end{equation}
and where
we have perturbations in the Hamiltonian-parameter $\lambda_1=\lambda_0+\delta\lambda$ and in temperature
$\beta_0=1/k_B T_0$, $\beta_1=1/k_B(T_0+\delta T)$. In the following we set Boltzmann's constant $k_B$ to unity.
It can be checked that when both temperatures $T_0$ and $T_1$ decrease to zero, the mixed-state fidelity 
reduces to the ground-state fidelity
$|\left\langle \phi_{GS}(\lambda)|\phi_{GS}(\lambda+\delta \lambda)\right\rangle|$, where $\left\vert \phi_{GS}(\lambda)\right\rangle$ is the ground state of Hamiltonian
$H(\lambda)$ for a particular value of the controlling parameter $\lambda$ .

When $\delta\lambda=0$, we define the temperature fidelity
$\mathcal{F}_{\beta}(\beta_0,\beta_1,\lambda)\equiv \mathcal{F}(\beta_0,\lambda;\beta_1,\lambda)$, which
 simplifies to
\begin{equation}
 \mathcal{F}_{\beta}(\beta_0,\beta_1,\lambda)=\frac{Z (\frac{\beta_0+\beta_1}{2}, \lambda)}
 {\sqrt{Z(\beta_0, \lambda) Z(\beta_1, \lambda)}}.\label{2}
\end{equation}
It can be further proved (see Appendix A) that  for small perturbations
$\delta T/T \ll 1$ \cite {zanardi3,gu}
\begin{equation}
 \mathcal{F}_{\beta}(\beta_0,\beta_1,\lambda)\approx e^{-\frac{(\delta\beta)^2}{8\beta^2}C_v},\label{3}
\end{equation}
where $C_v=-T\partial^2F/\partial T^2$ is the specific heat at constant field 
obtained from the free energy $F$ of the system.

When $\delta T=0$, we can define 
$\mathcal{F}_{\lambda}(\beta,\lambda_0,\lambda_1)\equiv \mathcal{F}(\beta,\lambda_0;\beta,\lambda_1)$, which
can be approximated as (see Appendix B)
\begin{equation}
\mathcal{F}_{\lambda}(\beta,\lambda_0,\lambda_1)
\approx\frac{Z(\beta, \frac{\lambda_0+\lambda_1}{2})}{\sqrt{Z(\beta,\lambda_0)Z(\beta,\lambda_1)}}.\label{4}
\end{equation}
The approximation in Eq. (\ref{4}) is due to the fact that in general $H(\lambda_0)$ and $H(\lambda_1)$ 
do not commute with each other, and is valid only for high temperatures 
such that  $\beta^3  \delta\lambda^3  \ll1$.
From Eq. (\ref{4}), and using arguments similar to those used for Eq. (\ref{3}), 
it can be shown that \cite{gu} (see Appendix C)
\begin{equation}
 \mathcal{F}_{\lambda}(\beta,\lambda_0,\lambda_1) \approx e^{-\frac{\beta(\delta\lambda)^2}{8}\chi},\label{5}
\end{equation}
where $\chi=- \partial^2F/ \partial \lambda^2$ is the susceptibility related to an external field
of strength $\lambda=(\lambda_0+\lambda_1)/2$. 

We see then from Eqs. (\ref{3}) and (\ref{5})
how the fidelity criterion for detecting a second-order phase transition \cite{zanardi1,zanardi2,zanardi3}
plays out for mixed-state fidelity: 
The minima of $ \mathcal{F}$ are associated with the singularities of the specific heat
and magnetic susceptibility.
More generally, as we will see below, 
$ \mathcal{F}$ inherits all non-analyticities of the free energy,
be them divergences or discontinuities in its second derivatives.
Therefore, it is reasonable to expect that fidelity can be used to study thermal phase transitions,
just like traditional criteria based on specific heat or susceptibilities.

It is interesting to note that from the above Eqs. (\ref{3}) and (\ref{5}) 
we can also obtain the so called perturbation-independent fidelity susceptibilities \cite{gu}
\begin{eqnarray}
\chi_{\beta}&\equiv\frac{-2 \ln { \mathcal{F}_{\beta}}}{(\delta\beta)^2}\approx\frac{1}{4\beta^2}C_v \label{6a}\\
\chi_{\lambda}&\equiv\frac{-2 \ln { \mathcal{F}_{\lambda}}}{(\delta\lambda)^2}\approx\frac{\beta}{4}\chi
\label{6b}
\end{eqnarray}
We would like to emphasize that Eq. (\ref{4}) and Eq. (\ref {6b}) hold approximately only for 
high temperatures, and are a bad approximation for low temperatures and especially for zero temperature, where quantum commutation relations are relevant.
We will discuss this point in Sec IV. Usually the calculation of 
$\mathcal {F}_{\lambda}$ is much more difficult than that of 
$\mathcal{F}_{\beta}$ due to the non-commutativity of $H(\lambda_0)$ and $H(\lambda_1)$.
In the following we will focus on the fidelity for a perturbation in temperature $\mathcal{F}_{\beta}$, 
and its application to second-order thermal phase transitions.
 
\section{Fidelity in second-order thermal phase transitions}

\begin{figure}[ht]
\begin{center}
\includegraphics[width=8cm, clip]{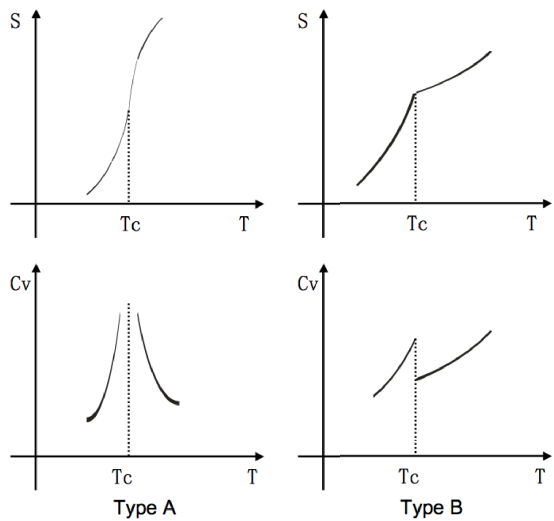}
\end{center}
\caption{Schematic diagrams of two types of second order phase transition.Type A corresponds to a 
divergence of second derivative of free energy, while Type B corresponds to a 
discontinuity (jump) of second derivative of free energy.  
In this example we plot the first and second derivative of free energy that 
correspond to entropy and specific heat respectively. 
Second-order phase transitions of Type A are also called $\lambda$ transitions. }
\label{fig0}
\end{figure}

Second order thermal phase transitions are characterized by non-analyticities in second derivatives of
the free energy (e.g. specific heat, susceptibility) with respect to
thermodynamic variables (temperature and external
magnetic fields). 
According to standard classification \cite{classification}, 
there are two types of non-analyticity that need to be considered: discontinuities and divergences 
(also known as $\lambda$ transition). 
For ordering purposes we call the associated transitions {\em type A} and {\em type B}, respectively, shown schematically in Fig. \ref{fig0}.

In the following we will discuss the behavior of fidelity near the critical points associated to these 
two types of thermal phase transitions.

\subsection{Type A: divergence of second-order derivative of free energy}
In this type of transition the specific heat at the critical point is much
larger than that at other points, and diverges in the thermodynamic limit. 
From Eq. (\ref {3}) we know that the critical point signaled by the maximum $C_v$ 
will correspond to a minimum of fidelity $ \mathcal{F}$.
Thus, for this type of systems the decay of fidelity as a function of the parameters 
(and with a fixed perturbation $\delta T$)
can be used to characterize accurately the phase boundaries. 
A good example of this situation is the two dimensional (2D) classical Ising model. 
This system is described
by the Hamiltonian $H=-J \sum_{\left\langle i,j \right\rangle}s_i s_j+\lambda \sum_i s_i$, where $\left\langle i,j \right\rangle$ means sum over nearest neighbor sites, $s_i=\pm1$, and
 $\lambda$ is the external magnetic field. 
 Onsager's famous solution \cite{huang} gives the partition function for zero 
 external magnetic field ($\lambda=0$),
 \begin{widetext}
 \begin{equation}
 Z(\beta, \lambda=0)=\exp{\left\{N \ln{[2\cosh(2\beta J)]}+\frac{N}{2\pi}\int_0^{\pi}{d \phi \ln {\left[\frac{1+\sqrt{1-K^2 (\sin{\phi})^2}}{2}\right]}}\right\}},\label{7}
 \end{equation}
 \end{widetext}
 where $K=2 \sinh{(2 \beta J)}/[\cosh(2 \beta J)]^2$. 
 By inserting this into Eq. (\ref{2}), we can obtain the fidelity  
 $ \left.\mathcal{F}_\beta(\beta_0,\beta_1,\lambda)\right|_{\lambda=0}$
 for the  2D classical Ising model (see Fig. 2a).
\begin{widetext}
\begin{figure}[ht]
\begin{center}
\includegraphics[width=7.5cm]{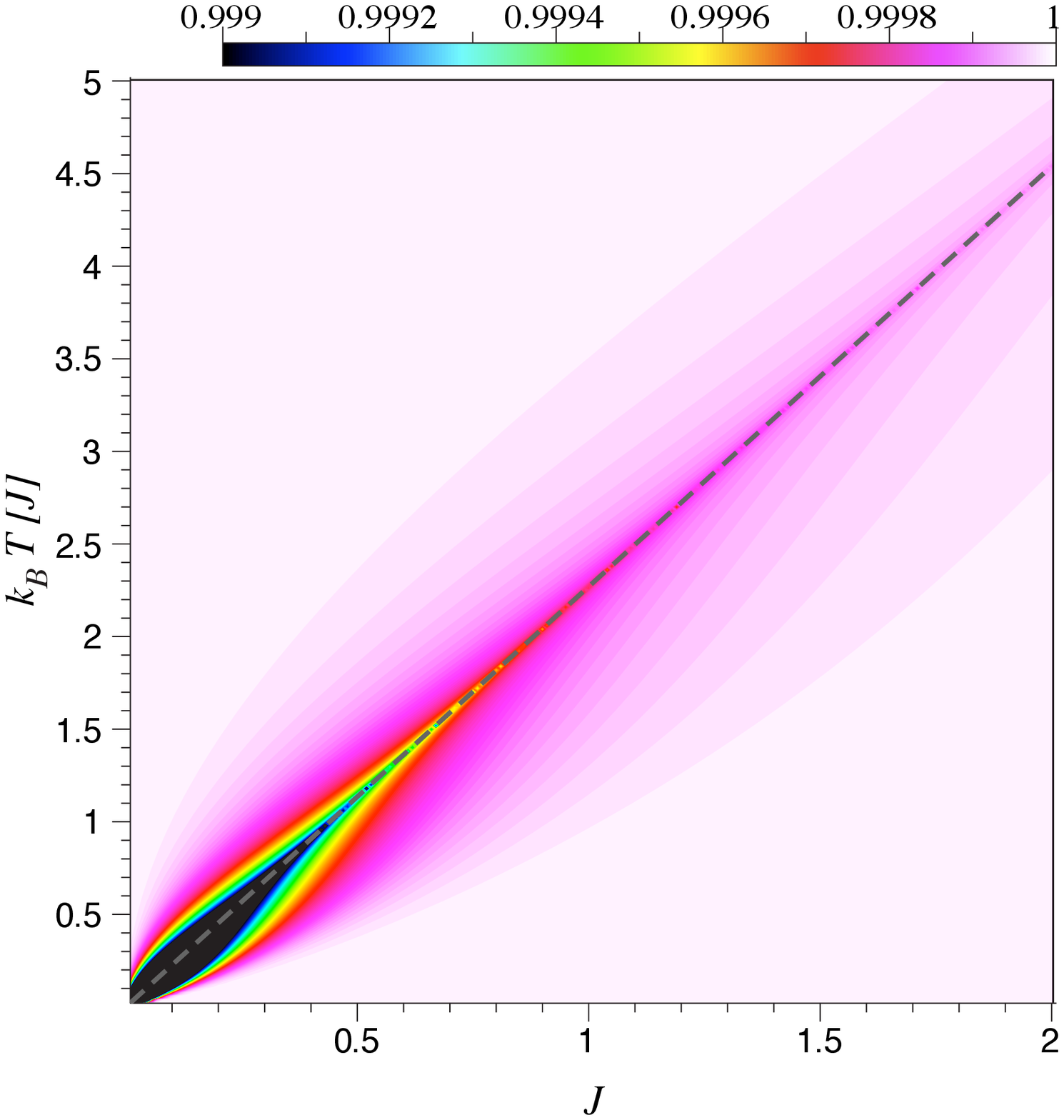}
\includegraphics[width=7.5cm]{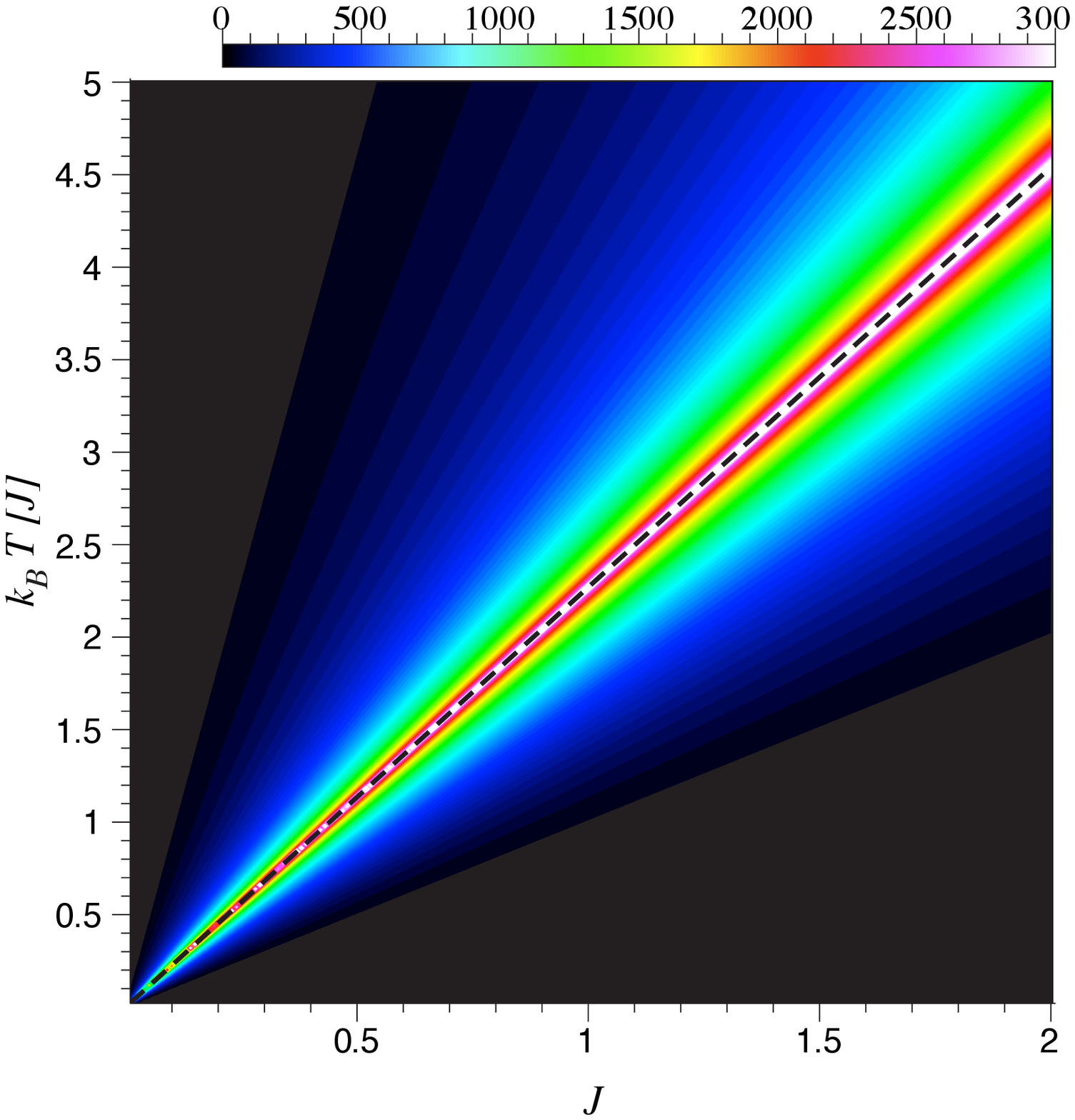}
\end{center}
\caption{Fidelity (left) and specific heat (right) 
of the 2D Ising model at zero external field. 
The critical temperature $T_c\approx2.27J$,
indicated by the dashed line, is clearly signaled by the minimum of fidelity. 
Because the specific heat diverges at the 
critical point, the 2D Ising model corresponds to a
Type A second-order phase transition.}
\label{fig1}
\end{figure}
\end{widetext}
The minimum of fidelity agrees well with the analytical result 
 for the critical temperature $T_c\approx2.27J$
 (see for comparison the specific heat on the right panel).
Since fidelity decays only on the critical lines, 
 we conclude then that its minimum is a good indicator of criticality in this type 
 of transitions. 
 Nevertheless, we would like to point out that the decay of 
 fidelity at the critical points becomes less drastic for higher temperatures. 
 This is because thermal fluctuations tend to wash out the information about 
 phase transitions encoded in the fidelity, an effect we will discuss in more detail in Section IV.
 
 \subsection{Type B: discontinuity of second-order derivatives of free energy}
\begin{widetext}
\begin{figure}[ht]
\begin{center}
\includegraphics[width=7.5cm]{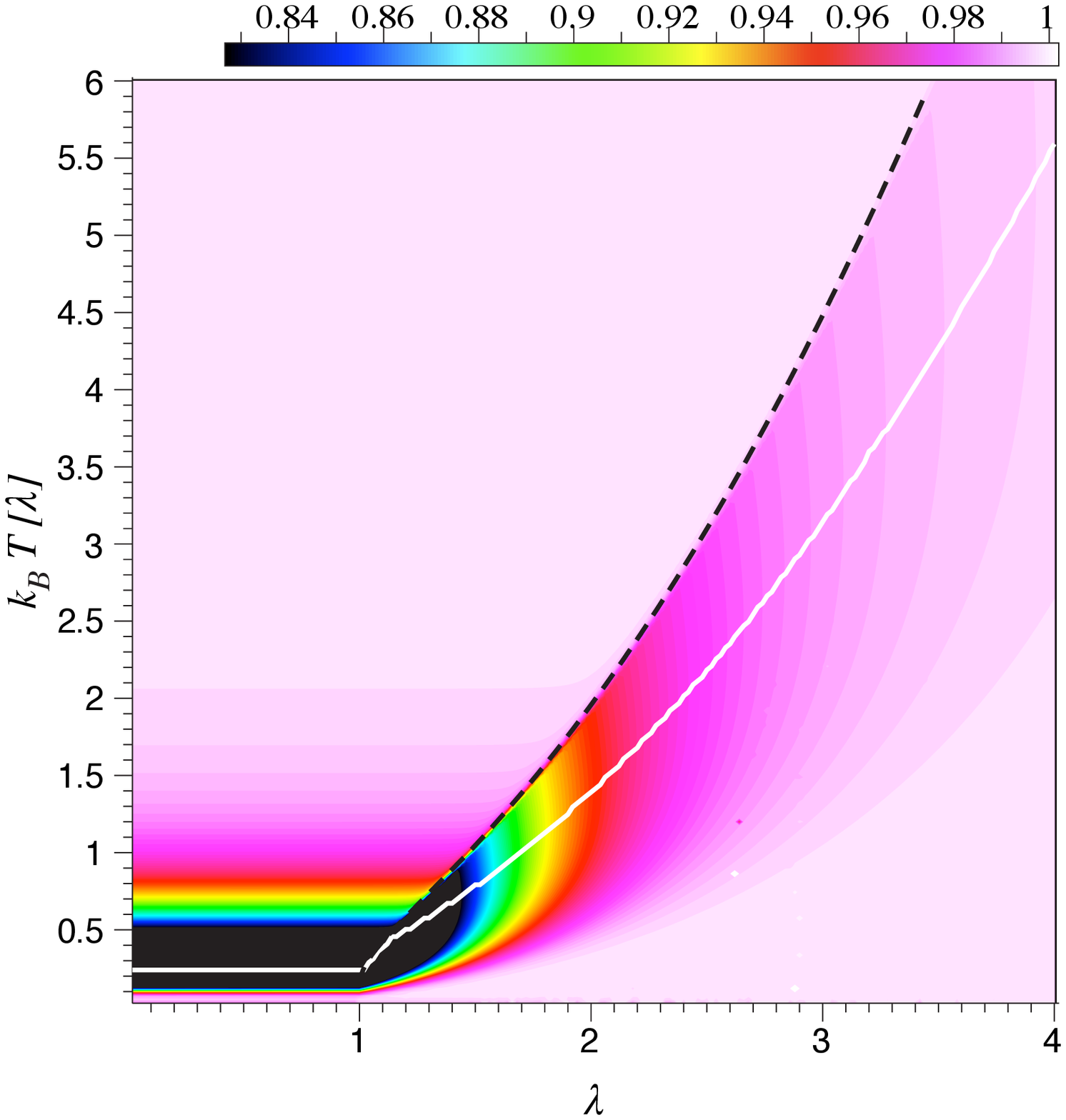}
\includegraphics[width=7.5cm]{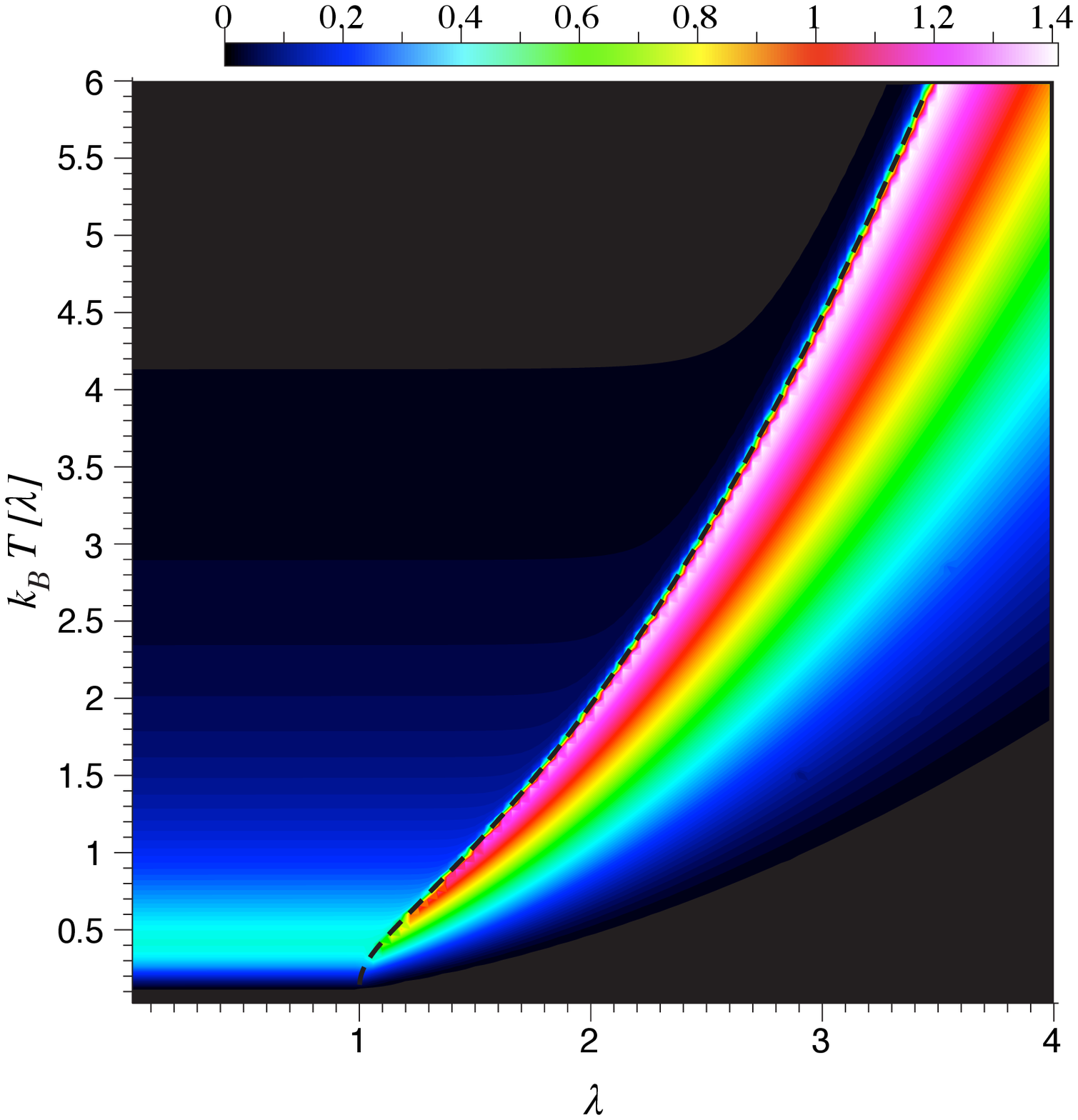}
\end{center}
\caption{Fidelity (left) and specific heat (right) of the 
Dicke model for $\omega=\omega_0$. On the left plot, the minimum of fidelity 
for fixed $\lambda$ is indicated with a white solid line.
The thermal phase transition line (dashed line in both plots) 
in the superradiant phase ($\lambda >1$) is indicated by a discontinuity in
specific heat and, accordingly, also in fidelity. 
The decay of fidelity in the normal phase ($0<\lambda<1$) is due to a
crossover instead of a thermal phase transition. The jump instead of a divergence in the specific heat indicates that the thermal phase transition in Dicke model belongs to a Type B 
second-order phase transition.
}
\label{fig2}
\end{figure}
\end{widetext}

A common type of transitions is characterized by a discontinuity or jump of second order derivatives of the free energy at the critical point. 
This is the case for instance in systems described by a simple Landau-Guinzburg theory \cite{classification}.
In such systems fidelity will not present in general a minimum at the critical points, but somewhere else in the phase diagram. 
A good example of these type of transitions is the Dicke
model, a collection of $N$ two-level atoms interacting with a single bosonic mode 
via a dipole interaction with an atom-field coupling
strength $\lambda$ \cite {dicke}. The Hamiltonian of the Dicke model can be written as
\begin{equation}
H=\omega_0J_z+\omega a^{\dagger}a+\frac{\lambda}{\sqrt{N}}(a^{\dagger}+a)(J_{+}+J_{-}),\label{8}
\end{equation}
where $a$ and $a^\dagger$ are annihilation and creation operators of the bosonic mode,
$J_z$, $J_+$, and $J_{-}$ are angular momentum operators of the total spin of the system,
$\omega$ and $\omega_0$ are the natural frequencies of the decoupled system, and $\lambda$ is the
spin-boson interaction strength.
Hamiltonian (\ref{8}) exhibits both a second-order thermal phase transition \cite{emary} 
and a quantum phase transition \cite {lieb, wang}, which has been studied
using ground-state fidelity \cite{zanardi1}. Here we will study the phase diagram of the Dicke model
at finite temperatures using mixed-state fidelity, Eq. (\ref{2}).
The exact partition
function of the Dicke model under the rotating wave approximation (RWA) is \cite{wang}
\begin{equation}
Z=2\int_{0}^{\infty} dr r e^{-\beta r^2} \left[2 \cosh\left(\frac{\beta \omega_0}{2 \omega}\sqrt{1+\frac{4\lambda^2 r^2\omega^2}{N \omega_0^2}}\right)\right]^N.\label{9}
\end{equation}
From this partition function one can obtain that there is a second-order phase transition
for $\lambda\ge1$ at a critical temperature
\cite{lieb,wang} 
\begin{equation}
\frac{1}{k_B T_c}=\frac{2\omega}{\omega_0}\tanh^{-1}\left(\frac{\omega_0}{\omega \lambda^2}\right).\label{10}
\end{equation}
From this partition function, Eq. (\ref{9}), 
we obtain the fidelity and the specific heat of the Dicke model (Fig. \ref{fig2}).
There are three aspects to highlight from the fidelity and 
specific heat shown in Fig. 3.
First, in the region where there is a thermal phase transition, $\lambda\ge1$, the minimum
of fidelity does not coincide with the phase boundary line.
This is easily attributable to the absence of a divergence in the specific heat,
which by means of Eq. (\ref{3}) implies that the minimum of fidelity need not be correlated to 
the transition line.
Second, fidelity presents minima in the region $0<\lambda<1$, where the
system only has a crossover (as seen from the specific heat, Fig. 3b).
This is again explained by the relation between fidelity and specific heat, Eq. (\ref{3}):
all maxima of the specific heat (which may not necessarily imply a thermal phase transition) will become minima of the fidelity.
We will explore more of this point in the next section.
Third, even though the specific heat has a visible discontinuity  
at the critical points,
fidelity changes rather continuously across the phase boundary,
specially at high temperatures.
Here we find a surprise, since from Eq. (\ref{3}) we would expect fidelity to be discontinuous at the critical points too. 
However, as mentioned before, thermal fluctuations affect fidelity strongly,
and this discontinuity is washed out for high temperatures.
With these three observations combined, we see that fidelity is actually not a 
good indicator of criticality for type B phase transitions: 
it cannot correctly signal the critical points with its minima, and is not reliable
with discontinuities. Furthermore, as in the $0<\lambda<1$ region of the Dicke model, fidelity might identify simple crossovers as phase transitions.

\section{Crossovers and thermal fluctuations}

As discussed in the introduction, geometrical arguments about fidelity in critical
systems lead us to expect that fidelity will have a minimum at the critical transition points. 
We just saw that this should be extended at least to identify discontinuities 
in fidelity with type B phase transition points
(akin to the behavior of ground state fidelity in first order QPTs). 
In this section we explore the 
following question: is it possible
to use fidelity, a quantum information tool, to fully characterize a critical system
at non-zero temperature,
i.e. by properly identifying all transition points of the phase diagram?

\subsection{Crossovers vs thermal phase transitions}
The free energy of a system is analytic everywhere in the $\lambda-T$ plane except at phase
transition points. But, there are many ``normal'' systems without transitions, i.e. 
where free energy is analytic simply everywhere. 
Nevertheless, this does not exclude the possibility that at some points the specific heat can become very large,
e.g. at the so called crossover points \cite{sachdev}. 
In fact, type A transitions in finite systems look like crossovers 
that become
divergences only at the thermodynamic limit.
Because of the relation between fidelity and specific heat, Eq. (\ref{3}), we expect that
fidelity will also have a minimum at the crossover point. This, in principle, can be seen as another
feature of fidelity, i.e. that fidelity can also be used to characterize crossovers \cite{zanardi3}. 
However, we are interested in the different problem of detecting a phase transition using fidelity.

Let us consider the example of the 1D Transverse Ising Model 
(TIM) with Hamiltonian $H=-J\sum_{i=1}^N(\sigma_i^z\sigma_{i+1}^z+\lambda\sigma_i^x)$. 
The partition function of the system is \cite {free energy}
\begin{equation}
\begin{split}
Z=2^N  \exp{ \left\{  \frac{N}{\pi}\int_{0}^{\pi}dk \ln\left[\cosh\left(\frac{2J\sqrt{1+\lambda^2-2\lambda \cos k}}{2k_BT}\right)\right] \right\} }
\end{split}
\end{equation}
\begin{widetext}
\begin{figure}[ht]
\begin{center}
\includegraphics[width=7.5cm, clip]{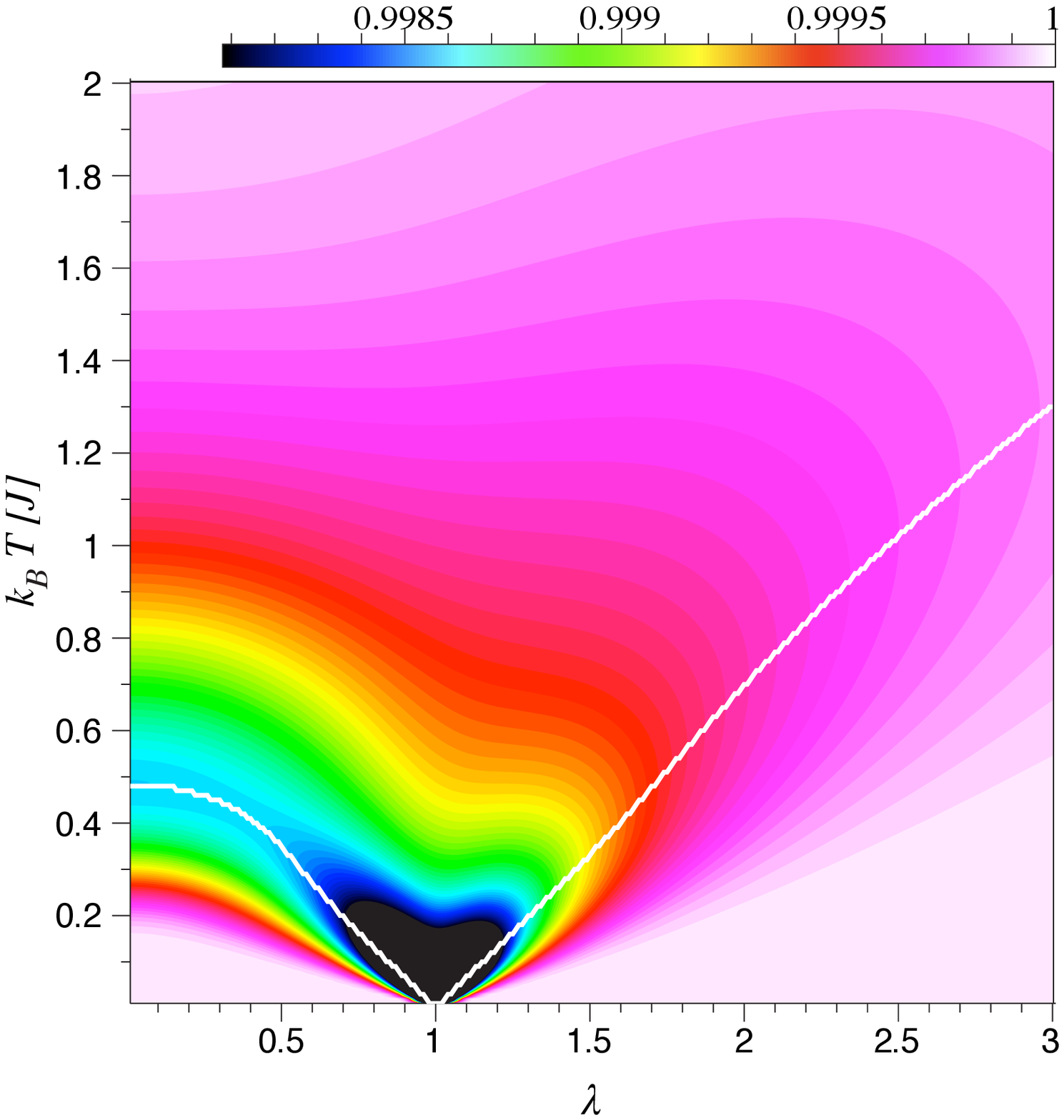}
\includegraphics[width=7.5cm, clip]{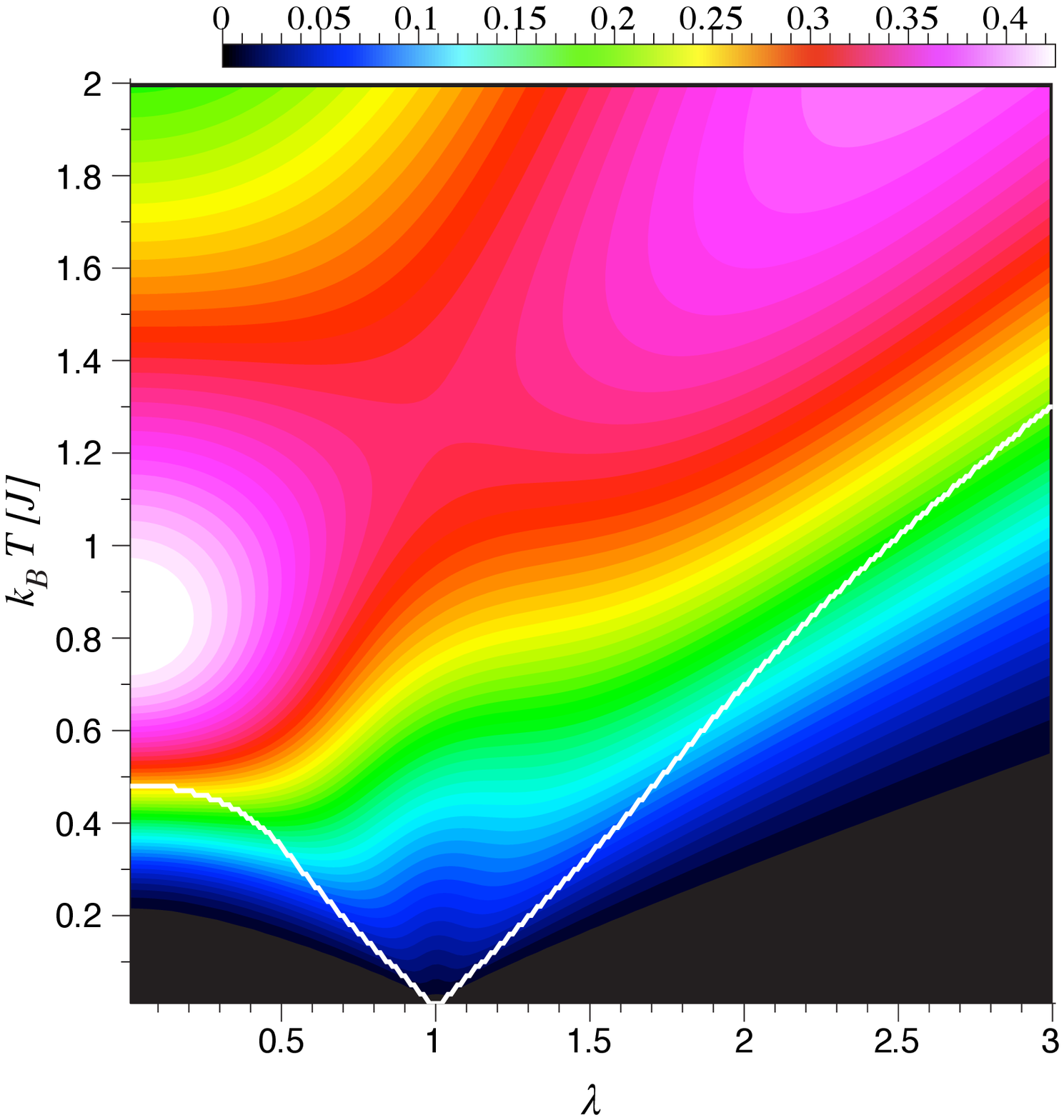}
\end{center}
\caption{Fidelity $\mathcal{F}_\beta(\beta_0,\beta_1,\lambda)$ (left) 
and specific heat (right) of 1D transverse Ising model with coupling $J$ in the 
$\lambda-T$ plane. 
The
white solid line indicates the minimum of fidelity.}
\label{fig6}
\end{figure}
\end{widetext}
We show a contour plot of fidelity for the 1D TIM in Fig. \ref {fig6} 
(a similar figure can be found in Ref. \cite{zanardi1}).
In this figure we see a minimum of fidelity following what appears as a phase transition line. 
We know, however, that this model does not have phase transitions for finite temperature
(one way to see this is to map the 1D TIM into a classical 2D Ising model, where the inverse
temperature is the effective {\em finite} 
size in the extra dimension of the classical system).
Thus, fidelity alone may not be able to distinguish simple crossovers from proper thermal phase
transitions. 
In order to make this distinction, we must resort to traditional statistical mechanics criteria -- like
the free energy and its derivatives. 
We show in Fig. 4 the specific heat for the 1D TIM in the $\lambda-T$ plane, which 
clearly does not have a divergence or a discontinuity.
A simple fidelity approach would also have led us to postulate a thermal phase transition
for the Dicke model for $0\le\lambda<1$ (see Fig. 3), which we know does not exist from the exact solution.
Even the applicability of fidelity to study crossovers is not clear: 
the ``crossover line" found with the minimum of fidelity line 
is different from $T_c\sim|1-\lambda|$ obtained in other discussions \cite{nonzero, scaling}

\subsection{Fidelity and thermal fluctuations} 
With fidelity arising from a quantum information approach, it is natural to question its 
behavior for moderate to high temperatures,
where quantum effects -- such as non-commutation of operators -- might be obscured.
Indeed, thermal fluctuations can wash out all information about phase
transitions characterized by ground-state fidelity \cite{zanardi2}. 
We already saw in the Dicke model and 2D Ising examples of previous sections that fidelity singularities become blurred for high temperatures, 
while the specific heat shows a singularity for all temperatures.
We refer again to figures 2, 3 and 4 for comparisons between $\mathcal{F}$ and $C_v$.
We see that the minimum of fidelity (or its discontinuity) becomes increasingly 
less prominent for larger temperatures,
eventually disappearing from the numerical precision.
On the contrary, specific heat is not influenced by thermal 
fluctuations and is a robust indicator of criticality up to very high temperatures.

Let us give a heuristic analysis of the influence of thermal fluctuations on mixed-state fidelity. 
The perturbation in temperature $\delta T$ can be expressed as
\begin{equation}
\delta\beta=\frac{1}{T}-\frac{1}{T+\delta T}=\frac{\delta T}{T(T+\delta T)}
\end{equation}
Hence, from Eq. (\ref{3}) we can write fidelity as
\begin{equation}
 \mathcal{F}_{\beta}\approx e^{-\frac{(\delta T)^2}{8T^2}C_v}, \label{13}
\end{equation}
From this equation we can see that when temperature increases, the effect on fidelity 
of the singularity of specific heat $C_v$ 
at critical points will be attenuated. 
For example, if the singularity in $C_v$ develops slowly with the size of a system, it might be 
very difficult
to detect it reliably using fidelity with finite size simulations.
It is important to highlight that the fidelity susceptibility  
\cite{gu} $\chi_{\beta}=C_v/(4\beta^2)=T^2 C_v/4$ 
will not be affected by thermal fluctuations at high temperature. 
Hence, even though fidelity itself may not be a good indicator of thermal phase transitions at high temperature, 
fidelity susceptibilities seem to be robust -- although this is just because
they are proportional to traditional quantities such as 
specific heat and susceptibility.

\section{Fidelity in the Lipkin-Meshkov-Glick model -- a case study}

For all the limitations we have discussed, fidelity decay or jump 
at the critical points is still
a necessary condition for a phase transition to exist. Therefore, in the cases where
fidelity is easier to compute than traditional observables from statistical mechanics
-- such as magnetic susceptibility and specific heat --, it can certainly
be used as a pre-criterion to explore the phase diagram of a system for potential 
phase transitions.
In order to test the predictive power of fidelity for thermal phase transitions,
we used it to study the phase diagram of the Lipkin-Meshkov-Glick (LMG) \cite {LMG} model of
$N$ globally coupled spins with an external magnetic field. The Hamiltonian of the
LMG model in units of the coupling energy is
\begin{equation}
H=-\frac{1}{N} \sum_{i<j} \left( \sigma_i^x \sigma_j^x + \gamma \sigma_i^y \sigma_j^y \right)
- \lambda \sum_i \sigma_i^z,
\label{LMG}
\end{equation}
where $\sigma^\alpha_i$, $\alpha=x,y,z$ are the Pauli matrices of the $i$-th spin, 
$\gamma$ is an anisotropy parameter, and $\lambda$ is
an applied external field.
We approached this problem without previous knowledge of its phase diagram, partly to test
the usefulness of fidelity, and partly (to be honest) out of ignorance.
In order to solve this model numerically we used a large spin $S=N/2$ representation, 
\begin{equation}
H=-\frac{1+\gamma}{N}\left(\vec{J}^2-J_z^2-\frac{N}{2}\right)-2\lambda J_z-\frac{1-\gamma}{2N}\left(J_{+}^2+J_{-}^2\right)
\label{lmg}
\end{equation}
where $J_{\alpha}=\sum_{i=1}^N \sigma_{i}^{\alpha}/2, \alpha=x, y, z$ is the total angular momentum operator. 
This is convenient because Hamiltonian (\ref{lmg}) does not mix subspaces with different
projection of the angular momentum, and one just has to diagonalize matrices of size $N\times N$.

\begin{figure}[ht]
\begin{center}
\includegraphics[width=10cm, clip]{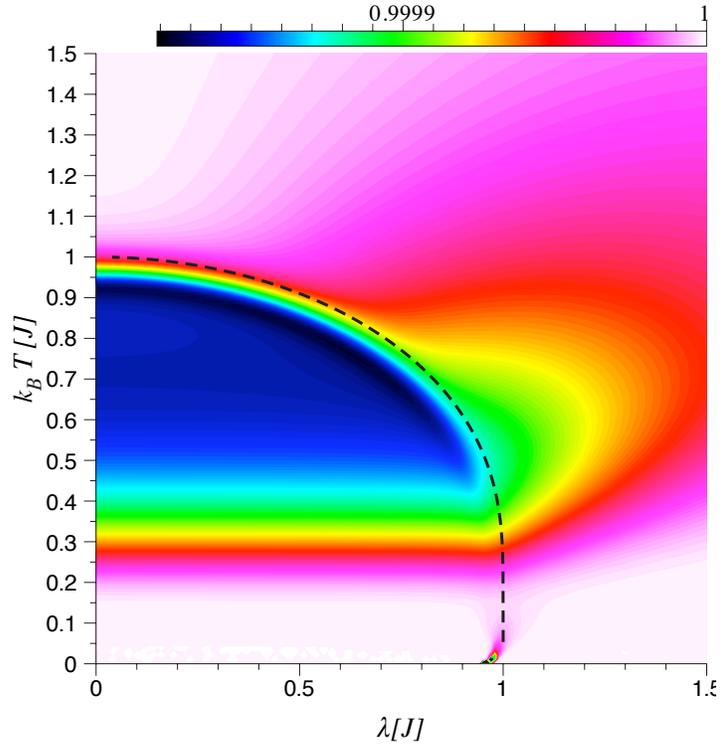}
\end{center}
\caption{Temperature fidelity of LMG model for $N=800$, and $\gamma=0.2$. 
Both the thermal phase transitions and quantum phase transition at $\lambda=1, T=0$ are indicated by the discontinuity and decay of fidelity respectively. 
Nevertheless, for the thermal phase transition, 
the discontinuity of fidelity deviates slightly 
from the phase boundary given by the mean-field result
(dashed line), although this is still within the deviation expected for a finite
size system. }
\label{fidelityLMG}
\end{figure}

Indeed, our fidelity studies detected something that appeared to be a thermal phase transition in the 
 $\lambda-T$ diagram (the anisotropy parameter $\gamma$ turns out to be not very important
as we will see shortly), see Fig. \ref{fidelityLMG}. We confirmed the existence of a thermal phase transition with further
numerical calculations of the specific heat and susceptibility, shown in Fig. 6, and by a mean field
calculation that we present here (we are not aware of such a calculation
for finite temperature in the literature).
\begin{figure}[ht]
\begin{center}
\includegraphics[width=10cm, clip]{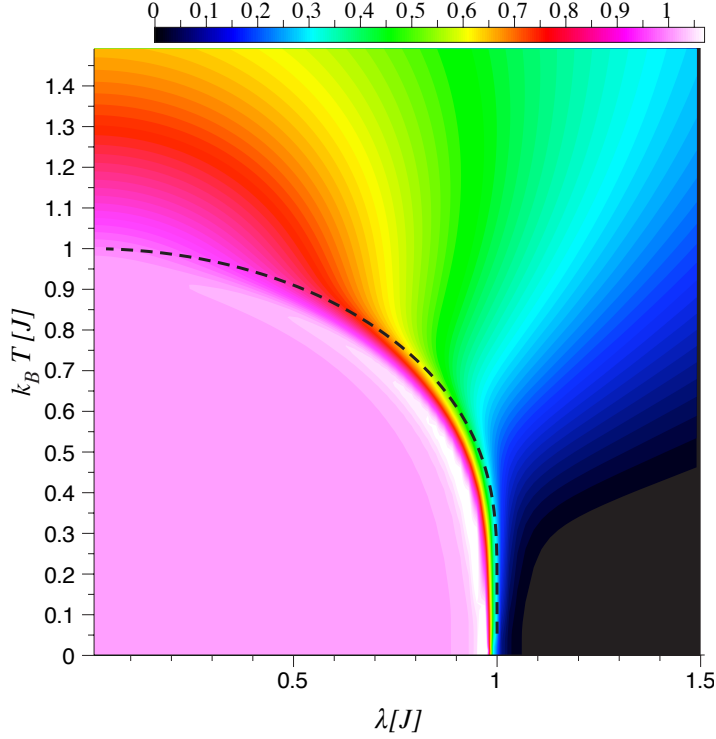}
\end{center}
\caption{Magnetic susceptibility $\chi$ of the LMG model for $N=800, \gamma=0.2$.
The phase boundary given by specific heat and susceptibility agrees well with that of mean-field result (shown in dashed line). Notice however that 
the discontinuity in $\chi$ for small $\lambda$ is less pronounced, given that
the phase boundary lies at almost constant temperature (the reverse
happens for the specific heat $C_v$ at low temperatures and $\lambda \approx 1$).}
\label{fig8}
\end{figure}
Under a mean-field approximation, the LMG Hamiltonian (\ref{LMG}) can be written as
\begin{equation}
\begin{split}
H=-&\frac{1}{2N}\sum_{i,j}[(\sigma_i^x-M_x)(\sigma_j^x-M_x) 
+M_x(\sigma_i^x+\sigma_j^x) \\
+&\gamma(\sigma_i^y-M_y)(\sigma_j^y-M_y)+\gamma M_y(\sigma_i^y+\sigma_j^y) \\
-& M_x^2-\gamma M_y^2]
-\lambda \sum_{i}\sigma_i^z
\end{split}\label {mean}
\end{equation} 
where $M_{\alpha}=\frac{1}{N}\left\langle\sum_{i=1}^N \sigma_i^{\alpha} \right\rangle, \alpha=x, y$ 
is the magnetization along the $\alpha$ direction. We will
see that this is also the order parameter of the phase transition of LMG model, in analogy with the
1D quantum XY model. 
The quadratic terms cancel out, and 
the Hamiltonian (\ref{mean}) is reduced to
\begin{equation}
H= \sum_{i=1}^N (-M_x\sigma_i^x- \gamma M_y \sigma_i^y-\lambda \sigma_i^z)+\frac{1}{2} M_x^2 +\frac{\gamma}{2} M_y^2 
\end{equation}
By now, the Hamiltonian is a sum of decoupled single-spin Hamiltonians that
can be diagonalized directly. The two eigenenergies are
\begin{equation}
\mathcal{E}_{\pm}=\pm\sqrt{M_x^2+\gamma^2 M_y^2+ \lambda^2}+\frac{1}{2}(M_x^2+\gamma^2 M_y^2),
\label{eigenenergy}
\end{equation}
and their corresponding eigenstates are
\begin{equation}
\begin{split}
\left\vert \mathcal{E}_{+}\right \rangle &=\frac{(M_x-i \gamma M_y)\left\vert \uparrow \right\rangle+(\mathcal{E}_{+} -\lambda)\left\vert \downarrow \right\rangle}{\sqrt{(M_x^2+\gamma^2 M_y^2)+(\mathcal{E}_{+} -\lambda)^2}},\\
\left\vert \mathcal{E}_{-}\right \rangle &=\frac{(M_x-i \gamma M_y)\left\vert \uparrow \right\rangle+(\mathcal{E}_{-} -\lambda)\left\vert \downarrow \right\rangle}{\sqrt{(M_x^2+\gamma^2 M_y^2)+(\mathcal{E}_{-} -\lambda)^2}},
\label{eigenstates}
\end{split}
\end{equation}
where $\left\vert \uparrow \right\rangle$ and $\left\vert \downarrow \right\rangle$ are eigenstates of $\sigma^z$. 
The self-consistent equations for the magnetization (order parameter) are
\begin{equation}
\begin{split}
M_x&=\left\langle\sigma_i^x \right\rangle=\frac{1}{z}e^{-\beta \mathcal {E}_{+}}\left\langle \mathcal{E}_{+} \right\vert \sigma_i^x \left\vert \mathcal{E}_{+}\right\rangle
+\frac{1}{z}e^{-\beta \mathcal {E}_{-}}\left\langle \mathcal{E}_{-} \right\vert \sigma_i^x \left\vert \mathcal{E}_{-}\right\rangle\\
M_y&=\left\langle\sigma_i^y \right\rangle=\frac{1}{z}e^{-\beta \mathcal {E}_{+}}\left\langle \mathcal{E}_{+} \right\vert \sigma_i^y \left\vert \mathcal{E}_{+}\right\rangle
+\frac{1}{z}e^{-\beta \mathcal {E}_{-}}\left\langle \mathcal{E}_{-} \right\vert \sigma_i^y \left\vert \mathcal{E}_{-}\right\rangle
\end{split} \label {selfconsistent}
\end{equation}
where $z=e^{-\beta \mathcal {E}_{+}}+e^{-\beta \mathcal {E}_{-}}$ is the partition function of the
mean field single-spin Hamiltonian. Combining Eqs. (\ref{eigenenergy}), (\ref{eigenstates}), and (\ref{selfconsistent}) we obtain the following two 
self-consistent equations
\begin{equation}
\begin{split}
M_x=&\frac{\tanh{[\beta\sqrt{\lambda^2+M_x^2+\gamma^2 M_y^2}]}}{\sqrt{\lambda^2+M_x^2+\gamma^2 M_y^2}}\times M_x\\
M_y=&\frac{\tanh{[\beta\sqrt{\lambda^2+M_x^2+\gamma^2 M_y^2}]}}{\sqrt{\lambda^2+M_x^2+\gamma^2 M_y^2}}\times \gamma M_y
\end{split}
\end{equation}
For $\gamma\neq 1$, the above two equations have nontrivial solutions only when either $M_x=0$ or $M_y=0$. 
The two parameters can be further determined by the condition of minimum free energy.
At absolute zero, the free energy equals the ground state energy $\mathcal{E}_{-}$. 
It is not difficult to find that when $\gamma<1$, $M_x\neq 0, M_y=0$ leads to the minimum ground state energy, while when
$\gamma>1$, $M_x= 0, M_y\neq 0$ leads to the minimum energy (and $M_x=M_y$ for
$\gamma=1$). 
For example, when $\gamma<1$, the self-consistent equation is reduced to
\begin{equation}
T=\frac{\sqrt{M_x^2+\lambda^2}}{\tanh^{-1}{\sqrt{M_x^2+\lambda^2}}}
\end{equation}
In the $\lambda-T$ plane, the phase boundary can be determined by setting the order parameter to be zero $M_x=0$. 
Then, the critical temperature
as a function of external magnetic field $\lambda$ is
\begin{equation}
T_c=\frac{\lambda}{\tanh^{-1}{\lambda}}, \ \ 0\leqslant\lambda\leqslant1
\end{equation}
This mean-field result agrees 
well with the phase boundary obtained by fidelity (Fig. 5) and traditional criteria,
 such as specific heat $C_v$ and magnetic
susceptibility $\chi$ (see Fig. 6), because the coordination number of the 
LMG model is $N-1$
-- i.e. it is is big enough to ensure the mean-field approximation is reliable. 

Thus, we have detected a phase transition with fidelity, which we then confirmed through an
analytical result. To the best of our knowledge, this is the first time these events
occur in this order, and lends support to our discussion above that fidelity is a good 
pre-criterion for testing phase boundaries.

\section{Discussion and Conclusions}
In the above Sec. III through Sec. V, we discussed the applicability of 
$\mathcal{F}_{\beta}$ to characterize thermal phase transitions, 
and indicated many of its limitations. Here we would like to further
consider the applications of $\mathcal{F}_{\lambda}$ and the ``quantum" (zero temperature) to ``classical" (nonzero temperature) transition of the system \cite{sachdev,vojta}. 
For high temperatures, 
statistical fluctuations dwarf quantum ones, and the importance of
uncertainty relations for the approximation in Eq. (\ref{4}) decreases.
In this case, the fidelity becomes a function of $\chi$ and the phase transition is classical \cite{sachdev,vojta}.
This means that with the increase of temperature, the fidelity criteria 
$\mathcal{F}_{\lambda}$ for QPT becomes equivalent to the susceptibility criteria for thermal phase transitions. 
Nevertheless, at low temperature, especially at zero temperature, the two criteria 
$\mathcal{F}_{\lambda}$ and $\chi$ differ dramatically due to the quantum and classical nature of the phase transitions.
This heuristic analysis agrees well with the result of Refs. \cite{sachdev,vojta}, 
that with the increase of the temperature 
the phase transition changes from ``quantum" to ``classical".

In summary, fidelity is a good tool to investigate quantum phase transitions, 
and has been extensively studied. 
However, when extending to finite-temperature thermal phase transition, it faces many limitations:
1) Fidelity decay occurs at both thermal phase transitions points and crossover lines, 
and fidelity alone can not distinguish between them. 
For this, we must fall back on traditional criteria such as the free energy and its derivatives.
2) For second-order phase transitions with a divergence in second derivatives of free energy (type A), 
drastic fidelity decay only occurs at critical points, and critical lines can be reliably indentified . 
However, for type B transitions -- with a discontinuity instead of a divergence --, 
fidelity decay occurs at many places besides critical points, and maximum decay of fidelity may 
not correspond to phase transition points. 
Fidelity itself might show a discontinuity, but it is easily visible only for low temperatures.
Hence, the standard fidelity-criterion for second-order thermal phase transitions 
is more applicable to type A than to type B thermal phase transitions.
3) In general, the fidelity approach is applicable to low temperature thermal phase transitions only. 
When the critical temperature is very high, fidelity may fail to signal the transition
because thermal fluctuations wash out all the relevant information 
encoded in fidelity. In comparison, traditional criteria based on free energy are not affected 
by thermal fluctuations and are good for any temperature. 

Before concluding this paper, we would like to point out that, despite its limitations for 
finite-temperature transitions, fidelity can still be a very useful {\em pre-criterion}
to detect thermal phase transitions, especially in systems where
we have no prior knowledge about its order parameter and symmetries, or even
topological thermal phase transitions without an order parameter and symmetry breaking. 
Because of its simple form, we can plot the fidelity of the system and then exclude the possibility of 
thermal phase transitions regimes without fidelity singularities. 
Afterwards, we can focus on suspect areas using free energy and traditional criteria
to distinguish crossovers from thermal phase transitions. 

\acknowledgements We would like to acknowledge 
Cristian Batista, Rishi Sharma and Michael Zwolak for stimulating discussions.


\appendix
\section{Specific heat and mixed-state fidelity}
We will see here the relation between fidelity with a temperature perturbation and
specific heat. From the standard definition $C_v=-T \partial^2 F/\partial T^2$, where $F$
is the free energy, and for a sufficiently small perturbation $\delta T/T \ll 1$, we can
approximate
\begin{equation}
\begin{split}
C_v(T)\simeq&-T \left[\frac{F(T+\delta T/2)+F(T-\delta T/2)-2 F(T)}{(\delta T/2)^2}\right] \\
\simeq & -\frac{2 T^2}{(\delta T/2)^2} \ln \frac{Z(T)}{\sqrt{Z(T+\delta T/2) Z(T-\delta T/2)}}
+ \frac{2 T}{\delta T} \ln \frac{Z(T+\delta T/2)}{Z(T-\delta T/2)}.
\end{split}
\end{equation}
Where we have used $F=-T \ln Z(T)$. Now, multiplying by 
$\delta \beta^2/\beta^2=\delta T^2/(T+\delta T)^2$, and keeping the lowest order terms in
$\delta T/T$,
\begin{equation}
\begin{split}
-\frac{(\delta \beta)^2}{8 \beta^2}C_v \simeq & 
\ln\frac{Z(\frac{\beta_0+\beta_1}{2})}{\sqrt{Z(\beta_0)Z(\beta_1)}},
\end{split}
\end{equation}
where $\beta_0=1/(T_0-\delta T/2)$, and $\beta_1=1/(T_0+\delta T/2)$. 
We thus obtain the relation between temperature 
fidelity $\mathcal{F}_{\beta}$ and specific heat
\begin{equation}
\mathcal{F}_{\beta}(\beta_0,\beta_1,\lambda)=\frac{Z(\frac{\beta_0+\beta_1}{2})}{\sqrt{Z(\beta_0)Z(\beta_1)}}
\simeq e^{-\frac{(\delta\beta)^2}{8\beta^2}C_v}.
\end{equation}

\section{non-commutative density matrix}
We look for a simplification of the perturbation in field fidelity,
\begin{equation}
\mathcal{F}_{\lambda}(\beta,\lambda_0,\lambda_1)=\mathrm{Tr}\sqrt{\sqrt{\rho_0}\rho_1\sqrt{\rho_0}}
\end{equation}
where $\rho_{\alpha}=\exp{(-\beta H(\lambda_{\alpha}))}/Z$. Usually, $H(\lambda_0)$ and $H(\lambda_1)$ do not commute with each other. However,
we can use the Trotter-Suzuki formula \cite{Suzuki} to approximate
\begin{equation}
\left\| \sqrt{\rho_0}\rho_1\sqrt{\rho_0}-\frac{e^{-\beta (H(\lambda_0)+\beta H(\lambda_1))}}{Z(\beta,\lambda_0)Z(\beta,\lambda_1)}
\right\| <  \beta^3 \Delta_2(H(\lambda_{0}),H(\lambda_{1}))
e^{\beta \left\| H(\lambda_{0}) \right\| + \beta \left\| H(\lambda_{1}) \right\| },
\label{eqSuzuki}
\end{equation}
where
\begin{equation}
\Delta_2(H(\lambda_{0}),H(\lambda_{1})) = \frac{1}{12}
\left( \left\| [[H(\lambda_{0}),H(\lambda_{1})],H(\lambda_{1})] \right\| +
\frac{1}{2}
\left\| [[H(\lambda_{0}),H(\lambda_{1})],H(\lambda_{0})] \right\|
\right).
\end{equation}
Thus, we have
\begin{equation}
\begin{split}
\mathcal{F}_{\lambda}(\beta,\lambda_0,\lambda_1)&\approx \frac{Z(\beta,\frac{\lambda_0+\lambda_1}{2})}{\sqrt{Z(\beta,\lambda_0)Z(\beta,\lambda_1)}}\\
&\approx \frac{Z(\beta,\lambda_0)}{\sqrt{Z(\beta,\lambda_0+\frac{\delta \lambda}{2})Z(\beta,\lambda_0-\frac{\delta \lambda}{2})}}
\end{split}
\end{equation}
which is Eq. (\ref{4}). The validity condition Eq. (\ref{eqSuzuki}) indicates that
at high temperature or small perturbation (typically $\beta^3 \delta \lambda^3\ll 1$), the fidelity criteria $\mathcal{F}_{\lambda}$ for QPT becomes equivalent to susceptibility criteria for thermal phase transition and thus the phase transition changes from ``quantum" to ``classical".

\section{Magnetic susceptibility and mixed-state fidelity}
Similar to Appendix A, we approximate the magnetic susceptibility 
\begin{equation}
\chi=-\frac{\partial^2 F}{\partial {\lambda}^2} \simeq \frac{F(\lambda_0+\frac{\delta \lambda}{2})+F(\lambda_0-\frac{\delta \lambda}{2})-2F(\lambda_0)}{(\delta \lambda/2)^2}\\
\end{equation}
Hence we have
\begin{equation}
\begin{split}
-\frac{\beta(\delta\lambda)^2}{8}\chi \simeq &-\frac{(\delta\lambda)^2}{8T} \frac{F(\lambda_0+\frac{\delta \lambda}{2})+F(\lambda_0-\frac{\delta \lambda}{2})-2F(\lambda_0)}{(\delta \lambda/2)^2}\\
\simeq &\frac{2 \ln{Z(\lambda_0)}-\ln{Z(\lambda_0+\frac{\delta \lambda}{2})}-\ln{Z(T_0-\frac{\delta \lambda}{2})}}{2}\\
\simeq &\ln{\frac{Z(\lambda_0)}{\sqrt{Z(\lambda_0+\delta \lambda/2)Z(\lambda_0-\delta \lambda/2)}}}
\end{split}
\end{equation}
From Appendix B, we obtain Eq. (\ref{5})
\begin{equation}
\mathcal{F}_{\lambda}(\beta,\lambda_0,\lambda_1)\approx \frac{Z(\lambda_0)}{\sqrt{Z(\lambda_0+\frac{\delta \lambda}{2})Z(\lambda_0-\frac{\delta \lambda}{2})}}
\simeq e^{-\frac{\beta(\delta\lambda)^2}{8}\chi}.
\end{equation}

\end{document}